\documentclass[aps,prl,amsmath,twocolumn,amssymb]{revtex4}%
\usepackage{graphicx}
\usepackage{pdfpages}
\usepackage{dcolumn}
\usepackage{bm}
\usepackage{amsmath}
\usepackage{fancybox}
\usepackage{amsfonts}
\usepackage{hyperref}
\usepackage{epstopdf}
\usepackage{appendix}
\usepackage{endnotes}
\usepackage{amssymb}%

\newcommand{\tdd}[2]{\frac{d #1}{d #2}}				
\newcommand{\Ham}{\mathcal{F}}					
\newcommand{\beq}{\begin{equation}}				
\newcommand{\eeq}{\end{equation}}					
\definecolor{darkgreen}{rgb}{.0,0.7,.0}

\date{\today}

\begin{document}
\title{Generation of single skyrmions by picosecond magnetic field pulses}

\author{Vegard Flovik}  \thanks{These authors contributed equally to this work}
\affiliation{Center for Quantum Spintronics, Department of Physics, Norwegian University of Science and Technology, NO-7491 Trondheim, Norway}

\author{Alireza Qaiumzadeh$^*$} \email{alireza.qaiumzadeh@ntnu.no}
\affiliation{Center for Quantum Spintronics, Department of Physics, Norwegian University of Science and Technology, NO-7491 Trondheim, Norway}

\author{Ashis K. Nandy}\thanks{These authors contributed equally to this work}
\affiliation{Peter Gr\"unberg Institut and  Institute for Advanced Simulation, Forschungszentrum J\"ulich and JARA, D-52425 J\"ulich, Germany}
\affiliation{Department of Physics and Astronomy, Uppsala University, P.O. Box 516, SE-75120 Uppsala, Sweden}

\author{Changhoon Heo}
\affiliation{Radboud University, Institute for Molecules and Materials, Heyendaalseweg 135, 6525 AJ Nijmegen, The Netherlands}
\author{Theo Rasing}
\affiliation{Radboud University, Institute for Molecules and Materials, Heyendaalseweg 135, 6525 AJ Nijmegen, The Netherlands}

\begin{abstract}
We numerically demonstrate an ultrafast method to create \textit{single} skyrmions in a \textit{collinear} ferromagnetic sample by applying a picosecond (effective) magnetic field pulse in the presence of Dzyaloshinskii-Moriya interaction. For small samples the applied magnetic field pulse could be either spatially uniform or nonuniform while for large samples a nonuniform and localized field is more effective. We examine the phase diagram of pulse width and amplitude for the nucleation. Our finding could ultimately be used to design future skyrmion-based devices.
\end{abstract}

\date{\today}
\maketitle
Skyrmions, nanoscale swirling magnetic structures, are topological solitons that might be utilized as magnetic bits in the next generation of compact memory devices \cite{Polyakov,Bogdanov_89,Bogdanov_94,Bogdanov_94_b,Bogdanov_99,Bogdanov_06, Bogdanov_11,Kiselev_11,Fert2013,Nagaosa-2013,Romming_2015,Hagemeister,Tomasello,Zhang15}, thanks to their small size and intrinsic stability. It is believed that nanoscale static skyrmions with a definite chirality and a fixed topological number $Q=\pm1$, \cite{suppmat} are stabilized in the presence of a chiral Dzyaloshinskii-Moriya interaction (DMI), so-called DMI stablized skyrmions \cite{Bogdanov_89,Bogdanov_94,Bogdanov_94_b,Bogdanov_99,Bogdanov_06, Bogdanov_11} while long-range dipole interactions stabilize bubble skyrmions \cite{Ezawa_10,Kiselev_2011,Klaui, Finazzi} with energetically degenerate chiralities and typically much larger diameters.

The possibility of a spontaneous skyrmion ground state in chiral magnets was first predicted by R\"{o}{\ss}ler \textit{et al.} \cite{Bogdanov_06}. The first observation of a skyrmion lattice, in the presence of a uniform magnetic field, was reported by M\"{u}hlbauer \textit{et al.} in 2009 in the chiral magnet MnSi \cite{Muhlbauer2009}. Since then, skyrmion lattices have been observed in various noncentrosymmetric materials and ultrathin helimagnets through different experimental techniques \cite{Muhlbauer2009,Yu2010,Heinze,Langner,Mehlin,Dussaux}. Recently, the generation of skyrmion lattices at room temperature was also demonstrated \cite{woo,Luchaire,Tokunaga}. Moreover a skyrmion lattice was observed in a Si-wafer-based multilayer system \cite{Schlenhoff}, making promises for future technology more realistic.

However, for applications, generation of isolated skyrmions is obviously more interesting than a lattice, as individual skyrmions can be used as bits of information \cite{Xichao,Zhou_rqce,Fert2013}. In case of a single skyrmion, the creation methods are distinct from the ones for skyrmion lattices and so far, few have been demonstrated, either numerically or experimentally. It was numerically shown that charge currents, spin-polarized currents, local heating, and spin waves can create isolated skyrmions in chiral ferromagnetic systems \cite{Tchoe,Lin,sampaio,iwasaki,koshibae, Liu_2015,Fujita}. All numerical simulations so far have predicted skyrmion nucleation in nanoseconds, and e.g. in Ref. [\onlinecite{Tchoe}] a circulating current is essential whereas in Ref. [\onlinecite{Lin}] an unrealistic damping parameter larger than 0.1 is needed. Romming \textit{et al.} observed the creation and annihilation of single skyrmions using a spin-polarized scanning tunneling microscope tip \cite{Romming13}, but to nucleate them an external magnetic field as well as very low temperature was needed.
On the other hand, bubble skyrmions stabilized by dipole interactions were observed in a ferrimagnetic thin film of TbFeCo using ultrashort single optical laser pulses \cite{Finazzi}, with a size of more than 100 nm. Creation of micron-sized synthetic skyrmions at room temperature has also been reported by current-induced spin-orbit torques (SOTs) \cite{skyrmionSOT}.
\begin{figure}[t]
 \centering
  \includegraphics[width=8.5cm]{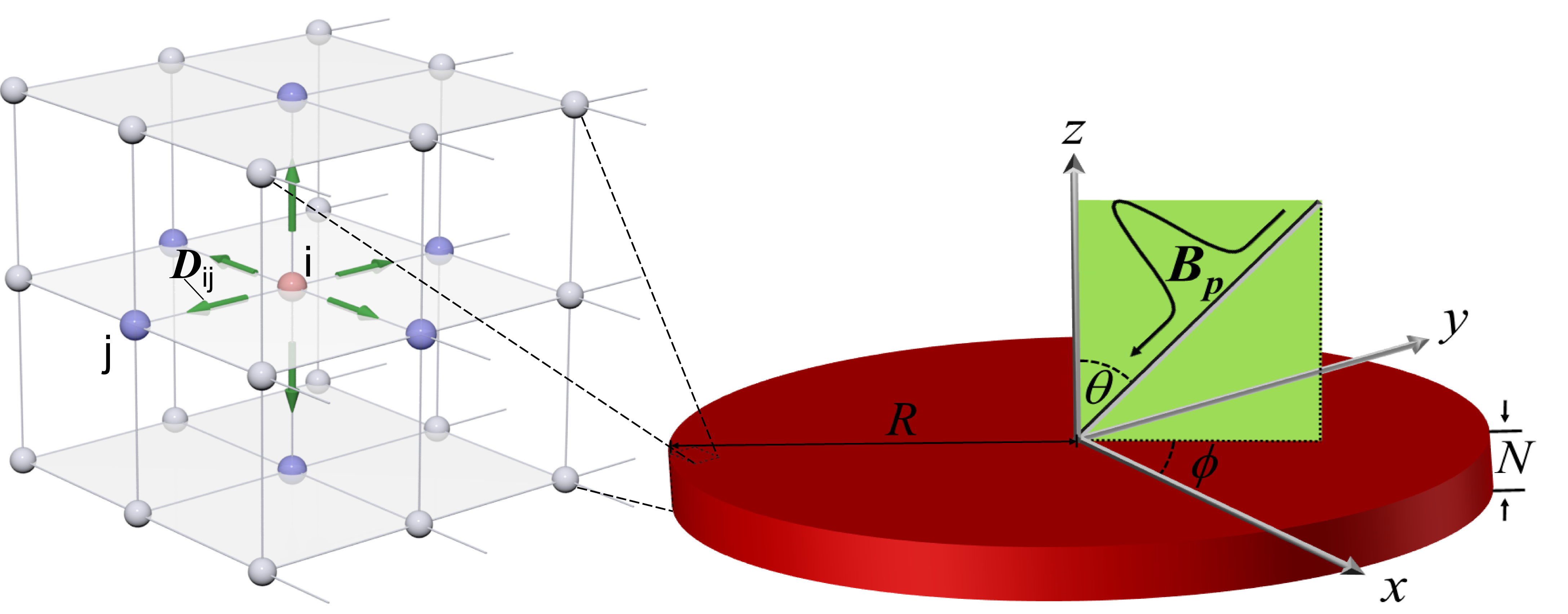}
  \caption{(Color online) Schematic representation of the simulated system. We consider a nanodisc of radius $R$ and thickness $N$ atomic layers, with an underlying square lattice structure. Direction of the Gaussian magnetic field pulse is defined by the polar angle $\theta$ and azimuthal angle $\phi$.}
 \label{fig_scheme}
\end{figure}
\begin{figure*}[t]
\centering
\includegraphics[scale=0.6]{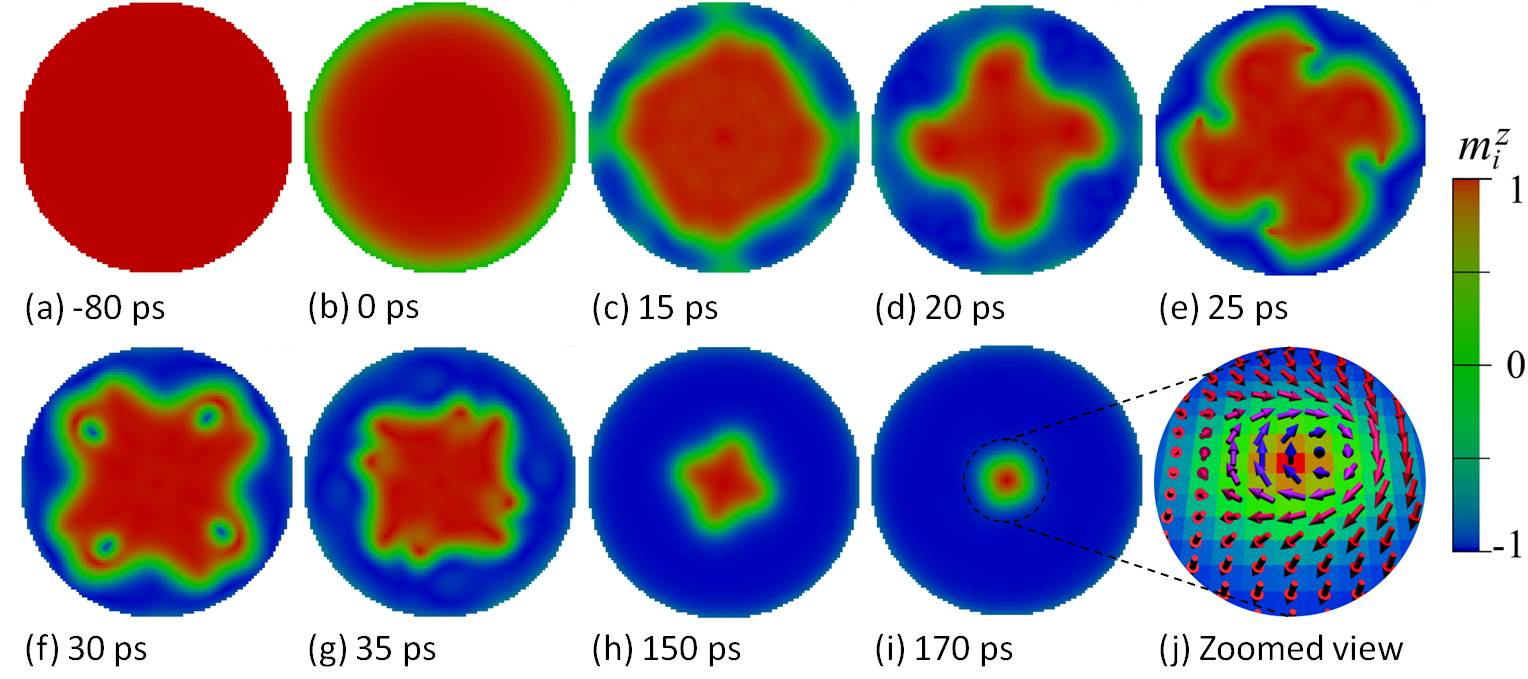}
\caption{(Color online) Time evolution snapshots of the magnetic spin configuration induced by a single 20 ps Gaussian magnetic field pulse normal to a circular nanodisc of $100a \times 3a$. (a) The initial state is a ferromagnetic state in the upward direction. (b) The maximum peak of a Gaussian magnetic field pulse arrives at $t=0$, and a domain with inplane spin direction emerges from the edge. (c) Remaining domain in the center has circular form, while the edge area shows reversed magnetization. (d)-(i) After passing of the pulse the shape of the center domain gradually changes from a circular to a cross shape, as a result of interference of spin waves. (j) Ultimately one chiral Bloch type skyrmion is stablized in the center of the nanodisc.}
\label{fig_sq}
\end{figure*}

In this Rapid Communication, we propose an efficient method for the nucleation of a skyrmion at a subnanosecond time scale in a uniform ferromagnetic sample using picosecond Gaussian magnetic field pulses, see Fig. \ref{fig_scheme}. Such fields can for example be created by applied laser pulses via the inverse Faraday effect \cite{IFE,IFE-Alireza1}.
For a large enough magnetic field pulse and for certain angles, spin waves are excited in the system and destabilize the ground state by exerting a magnonic SOT on the localized spins. Subsequently, an isolated skyrmion is dynamically nucleated in a metastable state in the presence of DMI. Successful nucleation of a skyrmion is observed, regardless of the sample geometry and size, using either spatially uniform or localized magnetic field pulses.

We use a classical model of localized magnetic spins. The total free energy of the system is given by,
\begin{align}
\Ham_{\mathrm{tot}}  =\Ham_{\mathrm{ex}} +\Ham_{\mathrm{DMI}} +\Ham_{\mathrm{ani}} +\Ham_\mathrm{Z},
\label{eq_modelHamiltonian}
\end{align}
where $\Ham_{\mathrm{ex}}=- J \sum_{i,j} \,  \bm{m}_i \cdot \bm{m}_{j}$ is the exchange interaction between neighboring local magnetic moments  $\bm{m}_i=\bm{\mu}_i/\mu_\textrm{s}$, and $J$ is the Heisenberg exchange constant. $ \Ham_{\mathrm{DMI}}=-\sum_{i,j} \bm{D}_{ij} \cdot  \bm{m}_i \times \bm{m}_{j}$ is the Dzyaloshinskii-Morya interaction where the DM vector $\bm{D}_{ij}$ for bulk DMI is defined as $\bm{D}_{ij}=D \bm{r}_{ij}$ and for interfacial DMI is $\bm{D}_{ij}=D (\hat{z}\times\bm{r}_{ij})$, with $D$ the strength of DMI, $\bm{r}_{ij}$ a unit vector pointing from site $i$ to site $j$, and $\hat{z}$ is the inversion symmetry breaking direction.
$ \Ham_{\mathrm{ani}}=- K \sum_{i} ({m}^z_i )^2$ is the anisotropy energy and $K$ is the out-of-plane uniaxial anisotropy constant. The coupling of an applied external magnetic field pulse $\bm{B}_\textrm{p}$, and the local spins is described by the Zeeman interaction  $\Ham_\mathrm{Z}=-\mu_\textrm{s}\sum_{i} \bm{m}_i \cdot \bm{B}_\textrm{p} $. Here we assume $\mu_\textrm{s}=2\mu_\textrm{B}$, where $\mu_\textrm{B}$ is the Bohr magneton.

We consider a thin film of magnetic atoms with open boundary conditions and a simple cubic lattice structure with lattice constant $a$, see Fig.~\ref{fig_scheme}. Our simulations show that, within the considered geometries, the effect of dipole interaction on the skyrmion nucleation is negligible, thus we ignore that. We assume a DMI strength less than the critical value $D<D_c=2\sqrt{J K}$, thus the ground state is collinear \cite{Stiles}.

The spin dynamics is described by the Landau-Lifschitz-Gilbert equation \cite{Lazaro98},
\begin{equation}
\label{SLLG}
\tdd{\bm{m}_i}{t}= \frac{\gamma}{(1+{\alpha}^2)\mu_s}
 \bm{m}_i \times ( \frac{\partial{\Ham}}{\partial{\bm{m}_i}}
                    +\alpha \bm{m}_i \times \frac{\partial{\Ham}}{\partial{\bm{m}_i}}),
\end{equation}
where $\gamma$ is the gyromagnetic ratio and $\alpha$ is the Gilbert damping parameter.
We have utilized a time dependent magnetic field pulse defined by a Gaussian function, $\bm{B}_\textrm{p}(t) = B_\textrm{0}(\bm{r}) \exp (-\frac{t^2 }{2 t^2_\textrm{w}}) \hat{e}_\textrm{B}$ applied in a direction $\hat{e}_\textrm{B}$. $B_\textrm{0}(\bm{r})$ and $t_\textrm{w}$ are the amplitude and the Gaussian width of the pulse \cite{FWHM}, respectively. $B_\textrm{0}(\bm{r})$ can be either uniform through the whole sample or localized with a Gaussian profile inside the domain. To calculate the topological number, $Q$, on a discrete lattice we use the Berg and L\"uscher formalism \cite{Berg81, suppmat}.

We used an atomistic spin dynamics simulation method implemented in the juSpinX code \cite{David} to study the temporal evolution of the magnetization. We also checked the results for larger samples, and investigated the effects of having a spatially localized and tilted magnetic field pulse by performing micromagnetic simulations using the simulation package MuMax3 \cite{mumax}.

In our atomistic simulations we consider a nanodisc of $100a \times 3a$ (diameter $\times $ thickness), and the time step is 1 fs. A typical simulation time is about 1 ns, which is long enough compared to a typical pulse width, $t_\textrm{w}$, of the order of tens of ps. The magnetic field is spatially uniform and covers the entire sample. The used material parameters for atomistic spin dynamics are the exchange stiffness J = 5 meV/atom and dimensionless parameters K/J = 0.1, D/J=0.16  and $\alpha$ = 0.05. In micromagnetic simulations we assumed a nanodisc of $500 \mathrm{nm} \times 1 \mathrm{nm}$ with a grid size of $2\mathrm{nm} \times 2\mathrm{nm} \times 1 \mathrm{nm}$, with material parameters for CoPt and FeGe thin films, see Supplemental Material \cite{suppmat}.

Figure~\ref{fig_sq} shows snapshot images of the temporal evolution of the magnetization in the nanodisc sample. The magnetic field pulse is applied at normal direction with an amplitude $B_0=3.5$ T and a pulse width $t_\mathrm{w}=20$ ps. We found similar dynamics in a rectangular geometry in the atomistic simulation. Also the micromagnetic simulations for large nanodisc sizes and applying tilted magnetic fields show similar temporal evaluation, see Supplemental Material \cite{suppmat}.

\begin{figure}[t]
\centering
\includegraphics[scale=0.7]{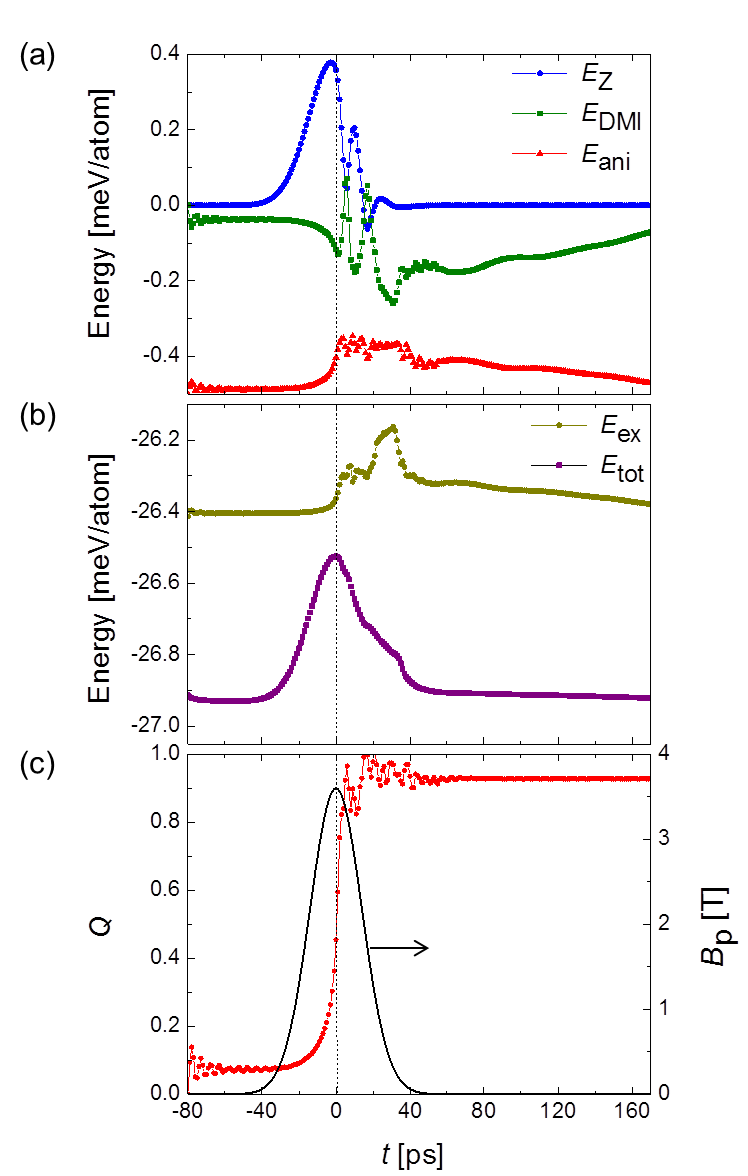}
  \caption{(Color online) Time dependent energy contributions and topological charge $Q$, in a circular nanodisc of $100a \times 3a$, during a spatially uniform magnetic field pulse with 3.5 T amplitude and 20 ps pulse width. (a) The Zeeman energy (blue dotted line) varies together with the changing of the magnetic field pulse until the pulse has completely passed the system around 50 ps. The green line indicates the variation of the DMI energy. The red line shows the change of the anisotropy energy. (b) Temporal evolution of the exchange energy and the total energy. (c) $Q$ changes from $Q\sim0$ to $Q\sim+1$ as the maximum peak of the Gaussian magnetic field pulse arrives at the system (vertical dashed line). Black line indicates the Gaussian magnetic field pulse.}
 \label{fig_sq_energy}
\end{figure}

When the peak of the incident pulse arrives at $t=0$, the sample becomes destabilized due to the excitation of spin waves and the subsequent magnonic SOT exerted on the local spins \cite{Manchon}. While all localized spins are collinear within the sample, at the edges the combination of DMI and finite size effects causes a tilting of the spin directions. Applying a uniform and perpendicular magnetic field pulse then excites spin waves at the edges. For the nucleation of a skyrmion, the amplitude of the pulse must be large enough to destabilize the uniform ferromagnetic state. The pulse amplitude turns completely off at about 40 ps. As the final outcome, a chiral vortex-like skyrmion in the center becomes stabilized and the topological charge of the systems becomes $Q\sim+1$ \cite{Q01}. The time to nucleate a skyrmion in this way is of the order of 200 ps and is much faster than in other proposed methods using currents and current pulses \cite{iwasaki,Lin}.
The formation of a domain with four fold symmetry in the center of the sample during the nucleation process, Figs.~\ref{fig_sq} (c)-(h), originates from the symmetry of the underlying lattice structure in the atomistic simulation, i.e. the square lattice, see also movie S1 in the Supplemental Material \cite{suppmat}. In the micromagnetic simulations for different material parameters also a four-fold symmetry appears, see movie S2 in the Supplemental Material \cite{suppmat}, but this is just an artifact of the finite difference discretization method.

The total energy of the system $E_\mathrm{tot}$, is the sum of the exchange energy $E_{\mathrm{ex}}$, the DMI energy $E_{\mathrm{DMI}}$, the anisotropy energy $E_{\mathrm{ani}}$, and the Zeeman energy $E_{\mathrm{Z}}$, Eq. (\ref{eq_modelHamiltonian}). Figures \ref{fig_sq_energy} (a and b) show the temporal evolution of each energy contribution resulting from the Gaussian magnetic field pulse $B_\mathrm{p}$. The separation into two figures is necessary due to the huge difference in energy scales between the exchange energy and the rest. Compared to the initial ferromagnetic state, the newly formed skyrmionic state is a metastable state \cite{Polyakov} with slightly higher total energy, but the DMI energy becomes lower. The changing energies due to $B_\mathrm{p}$ are accompanied by a changing $Q$ and, when all energies are stabilized at about 40 ps, $Q$ has eventually stabilized to one, Fig.~\ref{fig_sq_energy}(c).

\begin{figure}[t]
\centering
\includegraphics[scale=0.6]{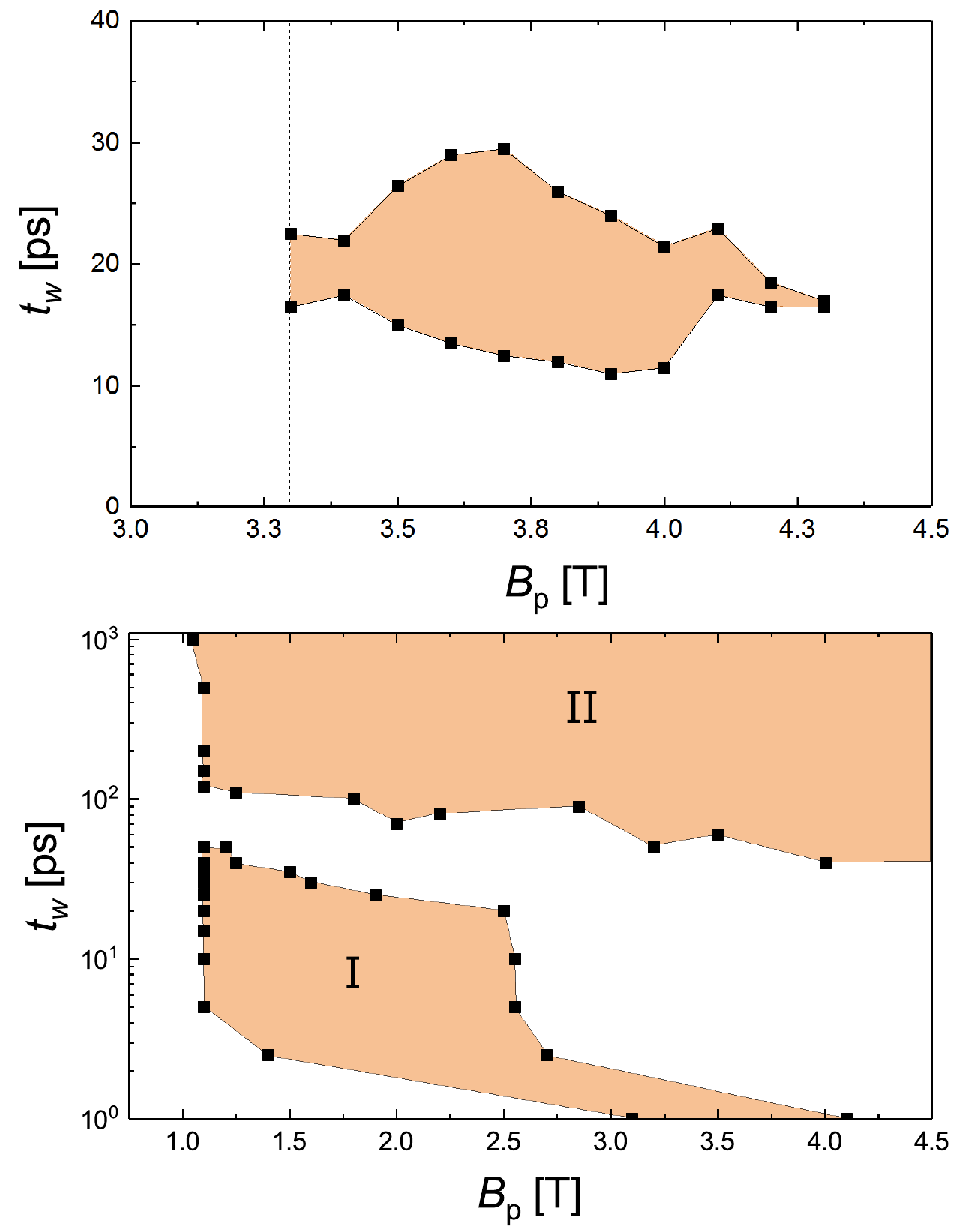}
\caption{(Color online) Phase diagram for the skyrmion nucleation induced by a magnetic field pulse: a) Atomistic simulation of single skyrmion generation by applying a uniform and perpendicular $\theta=0^\circ$ magnetic field pulse on a nanodisc with a size of $100a \times 3a$. Two vertical dashed lines indicate the minimum and maximum values of $B_{p}$ for the nucleation of a skyrmion.  b) Micromagnetic simulation of single (multiple) skyrmion generation, region I (II), by applying a localized and tilted magnetic field $\theta=45^\circ$ to a CoPt disk of 500 nm. See also movies in the Supplemental Material \cite{suppmat}}
 \label{fig_sq_pd}
\end{figure}

In Fig. \ref{fig_sq_pd} the phase diagrams of the nucleation of a skyrmion for both atomistic and micromagnetic simulations are shown. With the parameters chosen for the atomistic simulations the feasible area of skyrmion nucleation by applying a uniform and perpendicular magnetic field is given by a pulse strength from 3.3 T to 4.3 T and a pulse width from 10 ps to 30 ps for the circular shaped system, Fig. \ref{fig_sq_pd} a.  This phase diagram is material parameter dependent and for optimized material parameters one can reduce the filed amplitude to just few hundreds of mT \cite{suppmat}. To uncover whether this effect is only an edge effect, we did micromagnetic simulations for larger sample sizes in the presence of a localized and tilted applied magnetic field, with a temporal and spatial Gaussian profile. Skyrmion nucleation is even observed in larger samples by applying a spatially localized magnetic field pulse either far from or close to the sample edges. Because the ground state is collinear, to excite high enough spin wave amplitudes one needs to apply a tilted magnetic field pulse in such large systems. In Fig. \ref{fig_sq_pd}b the phase diagram of skyrmion nucleation for a CoPt nanodisc of $500$ nm is plotted. The applied magnetic field is tilted $45^\circ$, and localized at the center of the nanodisc with a spatial Gaussian width of $100$ nm.
We checked the skyrmion nucleation for two different material parameters, CoPt and FeGe, thin films with larger sizes, see Supplemental Material \cite{suppmat}. FeGe has bulk DMI while CoPt has interfacial DMI and consequently these two different symmetries lead to two different skyrmion structures, namely hedgehog (N\'{e}el type) skyrmions and chiral vortex-like (Bloch type) skyrmions, respectively. In both samples we found two nucleation regimes depending on the magnetic field duration. Short pulses create single skyrmions, indicated as regime I in Fig. \ref{fig_sq_pd}b, while longer pulses generate multiple skyrmions, indicated as regime II. Between these two regimes there is a gap in the phase space at which we could not find any stable skyrmion nucleation.
In larger systems, more stable skyrmions might be generated by longer magnetic field pulses. In small systems, while in the transient time regime, several skyrmion cores are created but because their distances are short, they repel each other and at the end, only one skyrmion is stabilized at the disk center.

These numerical results can be understood by the following minimal model. To overcome the barrier between a stable uniform ferromagnetic phase and a metastable skyrmionic phase, the ground state must be destabilized \cite{Lin,Shibata}. This is achieved by the creation of spin waves via the magnetic field pulse. Recently, it was shown that spin waves exert a torque on a uniform ferromagnet in the presence of DMI \cite{Manchon}. Considering a uniform ferromagnetic state in which the DMI is smaller than a certain threshold, the magnonic SOT is given by \cite{Manchon, Qaiumzadeh},
\begin{equation}
{\bm{T}}^{\mathrm{SOT}} = \tau_{\|}\bm{m} \times (\hat{z} \times \bm{j}_{m})+\tau_{\perp}\bm{m} \times (\hat{z} \times \bm{j}_{m})	\times \bm{m},
\end{equation}
where $\bm{j}_{m} = (a \gamma J/\mu_s) \bm{m} \times \nabla \bm{m}$ is the spin-wave current, and $\tau_{\parallel}$ and $\tau_{\perp}$ are phenomenological constants related to the DMI strength. We consider spin waves with frequency $\Omega$ and wavevector $\bm{k}$, as a deviation from a collinear state $\bm{m}= \hat{z}+\delta \bm{m} e^{i\Omega t+i{\bm{k}}\cdot{\bm{r}}}$. The spin-wave energy reads,
\begin{equation}
\hbar \Omega=  \frac{1-i\alpha}{1+{\alpha}^2}[a^2 J {k}^{2}+ \mu_s {B}_{\mathrm{p}}+ K+ a J (\tau_{\parallel}+i \tau_{\perp})({k}_{x}+ {k}_{y})].
\end{equation}
Finding the magnetic field in which the imaginary part of the dispersion relation becomes zero, $\mathrm{Im}[\Omega]=0$, and minimizing it with respect to the wavevector, we find a nonzero wavevector proportional to DMI strength,
${k}_{x}^{c} = {k}_{y}^{c} =  - (\tau_{\parallel}-{\tau_{\perp}/\alpha})/{(2 a)}$, that gives rise to a structural instability. In the absence of DMI the wavevector remains zero $\bm{k}=0$, which is equivalent to the initial uniform state.
Finally the critical field to create such instability in the presence of DMI is given by,
\begin{equation}
\mu_s {B}_{\mathrm{p}}^{c} = \frac{J}{2} (\tau_{\parallel}-\frac{\tau_{\perp}}{\alpha})^{2}-K.\label{critical-field}
\end{equation}
Equation (\ref{critical-field}) gives the critical magnetic field required to generate spin waves with a finite wavevector. This minimum magnetic field does not create a uniform mode $\bm{k}=0$, as for ordinary spin-wave excitations, but rather a spin wave with finite wavevector ($k_x=k_y\neq 0$) is excited by the magnetic field pulse. By applying a magnetic field ${B}_{\mathrm{p}} > {B}_\mathrm{p}^c$, the initial uniform ground state, $\bm{k}=0$, is destabilized, and transferred to the metastable skyrmion state $k_x=k_y\neq 0$.

The skyrmionic state in our simulations could not be stabilized in the absence of DMI. This process occurs for a certain range of material parameters \cite{heo}, e.g., the range of DMI, when $K/J \sim 0.1$, should be about $ 0.13 \leq D/J \leq 0.18$ in the atomistic trilayer circular sample. As the skyrmion diameter is proportional to $(D/J)^{-1}$, the skyrmion size is also limited. This range of DMI strength is experimentally accessible \cite{Romming_2015,Yang_2012,Nembach_2015}. Moreover, the damping value $\alpha$ is also important to create a skyrmion. We used here $\alpha = 0.05$, for both atomistic and micromagnetic simulations, which is a realistic value. For much higher damping than 0.05, the nucleation of a skyrmion cannot be stabilized due to the lack of sufficient spin wave propagation. This is in contrast with the current-induced single skyrmion case in which larger damping $\alpha>0.1$, is needed \cite{Lin}. Finally, within the parameters chosen in the case of a uniform magnetic field normal to the sample, the typical range of the amplitude of a pulse is 3 T to 4 T, with about 20 ps pulse width to nucleate an isolated skyrmion in the center of the nanodisc. On the other hand, in the case of applying a localized and tilted magnetic field, the threshold field amplitude is reduced to 0.5 T while the temporal field duration can be much longer.
In general a large enough external perturbation is needed to trigger the instability of the uniform initial state.

In summary, we showed that in the presence of either bulk or interfacial DMI and in a specific range of realistic material parameters, skyrmions can be generated by means of a magnetic field pulse of subnanosecond duration. Both spatially uniform and nonuniform localized pulses can create skyrmions in small and large samples. To achieve such magnetic field amplitudes and durations one can use the inverse Faraday effect arising from a circularly polarized laser pulse \cite{IFE,IFE-Alireza1}, while fields of around a Tesla with pulse duration of 100 ps can be found in present day write heads \cite{write-head}.
This method is faster than other methods which have been proposed yet and might be even used for skyrmion nucleation in ferromagnetic insulators as well as conducting ferromagnets. Our results may boost the building of a next-generation of magnetic devices utilizing the skyrmion state. Recently, it was shown that by using the interlayer exchange coupling, skyrmions can be stabilized in a wider range of parameters \cite{Ashis}. Thus we expect generation of skyrmions by using magnetic field pulse might be applicable to a large class of materials.

\section*{Acknowledgments}
Authors thank N. S. Kiselev, R. A. Duine, B. Dupe, M. V. Mostovoy and A. N. Bogdanov for fruitful discussions. This work was supported by the European union seventh framework program FP7-NMP-2011-SMALL-281043 (FEMTOSPIN), ERC Grant agreement No.~339813 (EXCHANGE), and the Dutch science foundation NWO-FOM 13PR3118.

\cleardoublepage


\section*{Supplementary material}

# Supplemental Material
## *Generation of single skyrmions by picosecond magnetic field pulses*

Vegard Flovik* and Alireza Qaiumzadeh*†  
*Center for Quantum Spintronics, Department of Physics,*  
*Norwegian University of Science and Technology, NO-7491 Trondheim, Norway*

Ashis K. Nandy*  
*Peter Grünberg Institut and Institute for Advanced Simulation,*  
*Forschungszentrum Jülich and JARA, D-52425 Jülich, Germany and*  
*Department of Physics and Astronomy, Uppsala University, P.O. Box 516, SE-75120 Uppsala, Sweden*

Changhoon Heo and Theo Rasing  
*Radboud University, Institute for Molecules and Materials,*  
*Heyendaalseweg 135, 6525 AJ Nijmegen, The Netherlands*  
(Dated: October 13, 2017)

**Topological charge on a discrete lattice**

The topological charge or skyrmion number, $Q$, is the number of times that local magnetic moments, $\boldsymbol{m}$, wind around the unit sphere,

$$Q = \frac{1}{4\pi} \int dx dy\, \boldsymbol{m} \cdot (\partial_x \boldsymbol{m} \times \partial_y \boldsymbol{m}). \tag{1}$$

DMI breaks the chiral symmetry of magnetic structures and, depend on the type of DMI, either skyrmions ($Q = +1$) or antiskyrmions ($Q = -1$) are energetically favourable [1]. But in the presence of dipolar interactions, bubble skyrmions with both signs and arbitrary integer topological charge number are energetically degenerate.

To describe the topological properties in atomistic simulations, we follow the Berg and Lüscher [2] definition of topological charge on a discrete lattice [2],

$$Q = \frac{1}{4\pi} \sum_l A_l, \tag{2}$$

where $A_l$ is the area of the spherical triangle with vertices $\boldsymbol{m}_1$, $\boldsymbol{m}_2$, and $\boldsymbol{m}_3$, and the summation runs over all elementary triangles $l$, defined on the square lattice. The area can be calculated via $\cos(A_l/2) = \left(1 + \sum_{m<n} \boldsymbol{m}_m \cdot \boldsymbol{m}_n\right)/\sqrt{2\overline{\Pi}_{m<n}(1 + \boldsymbol{m}_m \cdot \boldsymbol{m}_n)}$, where $m, n = \{1, 2, 3\}$. The sign of $A_l$ is determined as $\text{sign}[A_l] = \text{sign}[\boldsymbol{m}_1 \cdot (\boldsymbol{m}_2 \times \boldsymbol{m}_3)]$.

**Skyrmion nucleation phase diagram of FeGe thin films**

Bulk FeGe is a cubic helimagnetic material. But it was suggested that ultrathin films of FeGe layers might have an unaxial anisotropy [3]. An out-of-plane anisotropy can be also induced by distortion [4]. We applied a localized and tilted magnetic filed pulse with a temporal and spatial Gaussian profile to a FeGe nanodisc of 500 nm in diameter. Fig. S1 shows the phase diagram and the final skyrmion profile. FeGe has a bulk DMI and the generated skyrmion has a vortex-like profile. For the micromagnetic simulations we used the following parameters [3–6]:

$$M_{\text{sat}} = 384 \times 10^3 \text{Am}^{-1}, A_{\text{ex}} = 8.78 \times 10^{-12} \text{Jm}^{-1}, K = 3 \times 10^5 \text{Jm}^{-3}, D = 1.58 \times 10^{-3} \text{Jm}^{-2}, \alpha = 0.05,$$

where $M_{\text{sat}}$ is the saturation magnetization and $A_{\text{ex}}$ is the exchange stiffness constant. Also we consider a grid size of 2nm × 2nm × 1nm. The FeGe phase diagram is similar to the phase diagram of skyrmion generation in CoPt in the presence of a spatially and temporally non-uniform external magnetic field (see Fig. 4 b in the main text). By comparing Figs. 4 of the main text and Fig. S1a one can conclude that the threshold magnetic field is very sensitive to the material parameters and by optimizing these, one can reduce the magnetic field amplitude to just a few hundreds of mT.

2

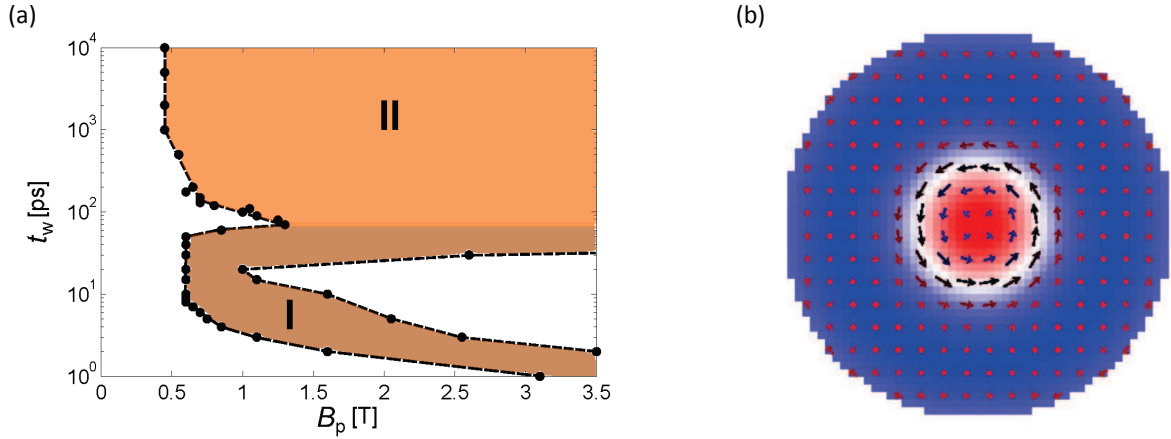

FIG. S1: (Color online) a) Phase diagram of single (regime I) and multiple (regime II) Skyrmion nucleation in a nanodic of FeGe with a diameter of 500 nm via applying a non-uniform and tilted $\theta = 45°$ magnetic field. b) Zoom view of the chiral Bloch type skyrmion generated in FeGe.

### Skyrmion nucleation phase diagram of CoPt bilayers

In Fig. 4b of the main text we represented the phase diagram of skyrmion generation in a CoPt nanodisc of 500 nm by applying a non-uniform and localized magnetic field with $\theta = 45°$. Here we present the phase diagram of a nanodisc of CoPt of 100 nm in diameter by applying a uniform and perpendicular ($\theta = 0°$) magnetic filed pulse with a temporal Gaussian profile. Fig. S2 shows the phase diagram and the final skyrmion profile. CoPt has an interfacial DMI and the generated skyrmion has a hedgehog (Néel) profile. For the micromagnetic simulations we used the following parameters [7, 8]:

$$M_{\text{sat}} = 580 \times 10^3 \text{Am}^{-1}, A_{\text{ex}} = 15 \times 10^{-12} \text{Jm}^{-1}, K = 0.8 \times 10^6 \text{Jm}^{-3}, D = 3.5 \times 10^{-3} \text{Jm}^{-2}, \alpha = 0.05.$$

Again we consider a grid size of 2nm × 2nm × 1nm. The final phase space is similar to the phase space of atomistic simulation of a nanodisc for the smaller size of $100a \times 3a$ (see Fig. 4a in the main text).

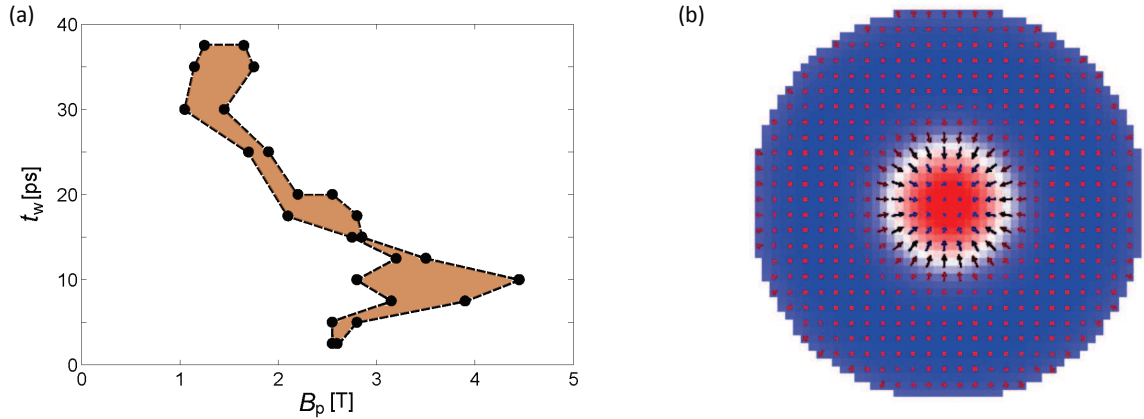

FIG. S2: (Color online) a) Phase diagram of single Skyrmion nucleation in a nanodic of CoPt with a diameter of 100 nm via applying a uniform and perpendicular magnetic field. b) Zoom view of the Hedgehog (Néel type) skyrmion generated in CoPt.

**Snapshots of skyrmion nucleation**

We provide four videos made from the snapshots of the out-of-plane magnetization component, see Fig. 2 of the main text:

1- File $S1.avi$ contains a movie of skyrmion generation by applying a spatially uniform magnetic field pulse on a nanodisc of $100a$ using atomistic simulation, where $a$ is the lattice constant.

2- File $S2.avi$ contains a movie of skyrmion generation by applying a spatially uniform magnetic field pulse on a nanodisc of $100nm$ using micromagnetic simulation.

3- File $S3.avi$ contains a movie of single skyrmion nucleation (regime I in Fig. 4b of the main text) by applying a localized and tilted magnetic field pulse on a CoPt nanodisc of 500 nm using micromagnetic simulation.

4- File $S4.avi$ contains a movie of multiple skyrmion nucleation (regime II in Fig. 4b of the main text) by applying a localized and tilted magnetic field pulse on a CoPt nanodisc of 500 nm using micromagnetic simulation. In the regime II, the direction of spins at the background is reversed.

**Acknowledgments**




\begin{thebibliography}{99}
\bibitem{FWHM}
In experiments, however, the full width at half maximum (FWHM) of the peak is often used and is given by  $\mathrm{FWHM}=2t_\mathrm{w}\sqrt{2\ln 2}\sim 2.35 t_\mathrm{w}$.
\bibitem{Q01}
Initial condition is a collinear state with $Q(t=-80)=0$. Having relaxed, the system obtains a finite small topological number thanks to the tilting of the spins at the boundaries. After applying a magnetic field pulse, a skyrmion is stabilized, but again because of the canted spins at the edges, the topological number of the system is less than unity.
\end{thebibliography}
\end{document}